# Analysis of Radon-Chain Decay Measurements: Evidence of Solar Influences and Inferences Concerning Solar Internal Structure and the Role of Neutrinos

P.A. Sturrock[1,], G. Steinitz[2], E. Fischbach[3]

P.A. Sturrock
   sturrock@stanford.edu
G. Steinitz
   Steinitz@gsi.gov.il
E. Fischbach
   Ephraim@physics.purdue.edu

[1] Kavli Institute for Particle Astrophysics and Cosmology and Center for Space Science and Astrophysics, Stanford University, Stanford, CA 94305-4060, USA
[2] Geological Survey of Israel, Jerusalem, 95501, Israel
[3] Department of Physics and Astronomy, Purdue University, West Lafayette, IN 47907, USA

## Abstract

An analysis of 85,000 measurements of gamma radiation associated with the decay of radon and its progeny in a sealed container located in the yard of the Geological Survey of Israel (GSI) in Jerusalem, between February 15, 2007 and November 7, 2016, reveals variations in both time of day and time of year with amplitudes of 4% and 2%, respectively. The phase of maximum of the annual oscillation occurs in June, suggestive of a galactic influence. Measurements made at midnight show strong evidence of an influence of solar rotation, but measurements made at noon do not. We find several pairs of oscillations with frequencies separated by 1 year$^{-1}$, indicative of an influence of rotation that is oblique with respect to the normal to the ecliptic, notably a pair at approximately 12.7 year$^{-1}$ and 13.7 year$^{-1}$ that match the synodic and sidereal rotation frequencies of the solar radiative zone as determined by helioseismology. Another notable pair (approximately 11.4 year$^{-1}$ and 12.4 year$^{-1}$) may correspond to an obliquely rotating inward extension of the radiative zone. We also find a triplet of oscillations with approximate frequencies 7.4 year$^{-1}$, 8.4 year$^{-1}$ and 9.4 year$^{-1}$ which, in view of the fact that the principal oscillation in Super-Kamiokande measurements is at 9.4 year$^{-1}$, may have their origin in an obliquely rotating core. We propose, as a hypothesis to be tested, that neutrinos can stimulate beta decays and that, when this occurs, the secondary products of the decay tend to travel in the same direction as the stimulating neutrino. The effective cross-section of this process is estimated to be of order $10^{-18}$ cm$^2$. The striking diurnal asymmetry appears to be attributable to a geometrical asymmetry in the experiment. Since we are here analyzing only gamma radiation, our analysis does not argue for or against an intrinsic variability of alpha-decays. Nighttime data show a number of curious "pulses" of duration 1 - 3 days.

Keywords Nuclear Physics – Solar structure

# 1 . Introduction

It is generally believed that all nuclear decay rates are constant, but there have for some time been hints that this may not be strictly correct. Emery (1972), in a review article, concluded that there is evidence that *screening…affects the rate of beta decay, whether electron or positron decays are considered*. Alburger, Harbottle and Norton (1986), following a four-year sequence of measurements at the Brookhaven National Laboratory (BNL) of the decay rate of $^{32}$Si, reported finding *small periodic annual deviations of the data points from an exponential decay curve [that were] of uncertain or*igin. Siegert, Schrader and Schotzig (1998), at the Physikalisch-Technische Bundesanstalt (PTB), reported the results of a 20-year study of the decays of $^{152}$Eu and $^{154}$Eu, using $^{226}$Ra as a standard, noting annual oscillations in the measured decay rates of both $^{152}$Eu and $^{226}$Ra. Falkenberg (2001), in his study of the beta decay of tritium, found evidence for an annual oscillation that appeared to be related to the annually changing distance of the Earth from the Sun, and suggested that this effect may be due to solar neutrinos.

The annual oscillations apparent in BNL and PTB data attracted the interest of Fischbach and Jenkins and their colleagues, who also suggested that solar neutrinos may be responsible for variations in beta-decay rates (Jenkins, Fischbach, Buncher, *et al.* 2009; Fischbach, Buncher, Gruenwald, *et al.* 2009). These articles drew the critical attention of Cooper (2009), to which Krause, Rogers, Fischbach, *et al.* (2012) responded; of Norman, Browne, Shugart *et al.* (2009), to which O'Keefe, Morreale, Lee, *et al.* (2013) responded; and of Semkow, Haines, Beach, *et al.* (2009), to which Jenkins, Mundy and Fischbach (2010) responded. The most recent critical articles are one by Kossert and Nahle (2014), to which we have responded (Sturrock, Steinitz, Fischbach, *et al.* 2016), and one by Pommé, Stroh, Paepen, *et al.* (2016), to which we are unable to respond since the relevant data are not available for analysis.

The early claims of variability of decay rates drew upon evidence for annual oscillations (Falkenberg, 2001; Jenkins, Fischbach, Buncher, *et al.,* 2009). However, such claims are open to the valid concern that annual oscillations may be caused by environmental variations (Semkow, Haines, Beach, *et al.,* 2009). To avoid this concern, we have for some time chosen to search for evidence of oscillations in a frequency range that can be associated with internal solar rotation, which we now take to be 9 – 13 year$^{-1}$ (Javorsek, Sturrock, Lasenby, *et al.,* 2010; Sturrock, Buncher, Fischbach, *et al.,* 2010a; Sturrock, Buncher, Fischbach, *et al.,* 2010b; Sturrock, Fischbach, Scargle, 2016; Sturrock, Steinitz, Fischbach, *et al.,* 2012; Sturrock, Steinitz, Fischbach, *et al.,* 2016). Our most recent article (Sturrock, Fischbach, Scargle, 2016) draws attention to a similarity between oscillations with frequencies in the range 9 – 13 year$^{-1}$ that are evident in the decay data acquired at BNL and in neutrino data acquired by the Super-Kamiokande Observatory, which we suggest may be attributed to influences of rotation in the solar interior.

To the best of our knowledge, the most extended nuclear-decay dataset now available is one currently being compiled by one of us (GS) at the Geological Survey of Israel (GSI) (Steinitz, Kotlarsky, Piatibratova, 2011). This experiment, which records 5 nuclear measurements (from 3 internal and 2 external detectors) and three environmental measurements every 15 minutes, has been in continuous operation since February 15, 2007. We here analyze 85,283 hourly measurements (up to November 7, 2016), made by a gamma-ray detector that is located vertically above a source of radioactivity ($^{238}$U) whose daughters decay into $^{222}$Rn, which has a half-life of 3.82 days. The layout of the experiment is shown in Figure 1 and the decay chain is shown in Figure 2.

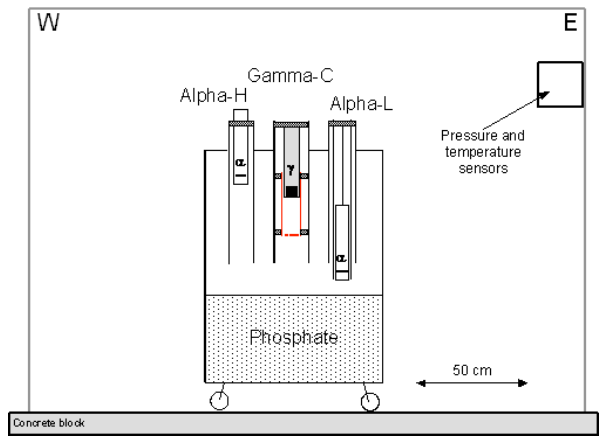

Figure 1. Experimental layout, showing the locations in the tank of the phosphorite (the source of $^{222}$Rn) and the three internal radiation detectors. This article analyzes measurements made with the gamma-C detector, which is located in a lead tube (thickness 0.5 cm). Pressure and temperature sensors, a data logger, and a continually loaded 12V battery are located in a separate enclosure as indicated in the figure. A full description is given in Steinitz *et al.*, 2011.

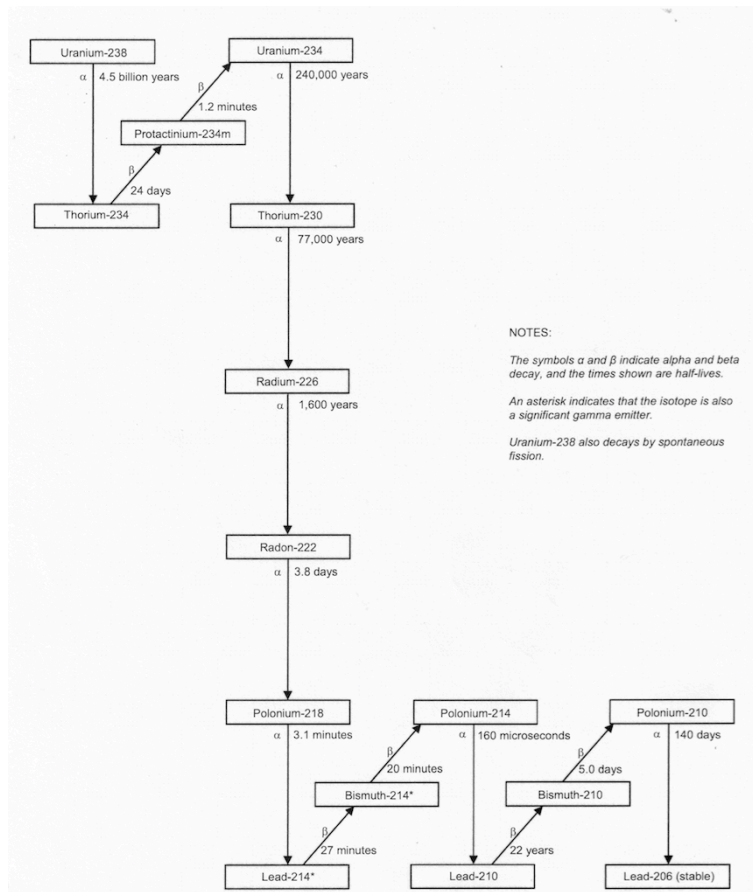

Figure 2. Decay chain of Uranium-238. (Courtesy of Argonne National Laboratory).

The gamma photons are produced by electrons that result from the beta-decays in the following four stages of the decay chain: $^{214}$Pb to $^{214}$Bi, $^{214}$Bi to $^{214}$Po; $^{210}$Pb to $^{210}$Bi, and $^{210}$Bi to $^{210}$Po. To the best of our knowledge, there is no evidence of *intrinsic* variability of the alpha-decay process (Parkhomov, 2010a, 2010b, 2011). However, measurements of the $^{214}$Po to $^{210}$Pb and $^{210}$Po to $^{206}$Pb alpha-decay rates must be expected to vary since these decays devolve from progenies of beta decays.

We review the data in Section 2 and present the results of power-spectrum analysis in Section 3. In Section 4, we draw attention to pairs of oscillations with frequencies separated by 1 year$^{-1}$, which we interpret as sidereal and synodic frequency pairs, such as one expects to be generated by a source such as the solar interior if it has a rotation axis that is oblique with respect to the normal to the ecliptic (Sturrock and Bai, 1992). Section 5 shows spectrograms that present power spectra as functions of time of day and, for comparison, similar spectrograms formed from environmental data (ambient temperature, ambient pressure and battery voltage). We find that there is no correspondence between spectrograms formed from the gamma measurements and those formed from the environmental measurements. We discuss these results in Section 6, comparing our results with the hypothesis that beta decays may be stimulated by neutrinos. Appendix A gives an estimate of the effective cross section of the neutrino process that we suggest may be responsible for beta-decay variability.

**2 . GSI Data**

For convenience of power spectrum analysis, we adopt a date format that does not involve leap years (Sturrock, Fischbach, Scargle, 2016). We first count dates in "neutrino days," for which January 1, 1970, is designated "neutrino day 1" ($t(ND) = 1$). We then convert dates to "neutrino years", denoted by $t(NY)$, as follows:

$$t(NY) = 1970 + t(ND)/365.2564 . \qquad (1)$$

Dates in neutrino years differ from true dates by less than one day.

We next detrend (to remove the exponential decay) and normalize the data as follows. From the times $t_r$ and count-rate measurements $y_r$, we form

$$x_n = y_n/z_v \text{ where } z_n = \exp(K - \kappa t_n) . \qquad (3)$$

and determine the values of $\kappa$ and $K$ that minimize $V$. The normalized and detrended values are then given by

$$x_n = y_n/z_v \text{ where } z_n = \exp(K - \kappa t_n) . \qquad (3)$$

The mean gamma-ray count rate is 762,000 counts per hour (212 per second). A plot of the residuals (in percent) of the normalized count-rate is shown as a function of time in Figure 3.

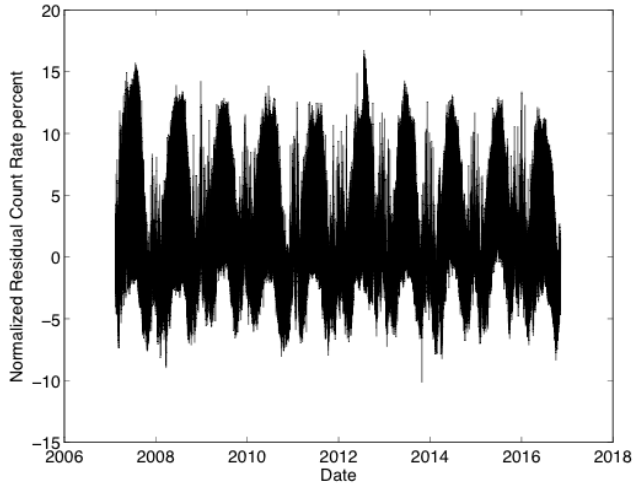

Fig. 3. Gamma measurements, normalized to mean value unity, percent deviation, as a function of time.

We see a complex pattern with a strong annual oscillation with an amplitude of about 10%. To examine the cause of this complexity, we show in Figure 4 a small section of the data (from 2007.125 to 2007.220). This plot shows that there is a strong *diurnal* variation in the measurements, with mean amplitude of approximately 5%.

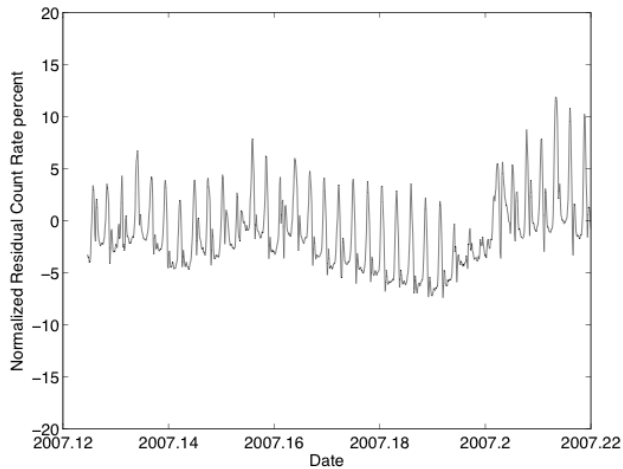

Fig. 4. Small section of gamma measurements, normalized to mean value unity, percent deviation, as a function of time. The sharp fluctuations represent variations on a daily scale.

In view of these diurnal oscillations, we determine the local (at Jerusalem) "hour of day" of each measurement. We sort the measurements by hour of day and then organize the data into 100 bins. For each bin, we determine the mean and the standard error of the mean. The result is shown in Figure 5. However, the standard error of the mean is so small that it is convenient to show curves that differ from the mean value by 10 times the standard error of the mean. We see that there is a sharp peak at about 12 hours (noon) and a smaller peak at about 20 hours.

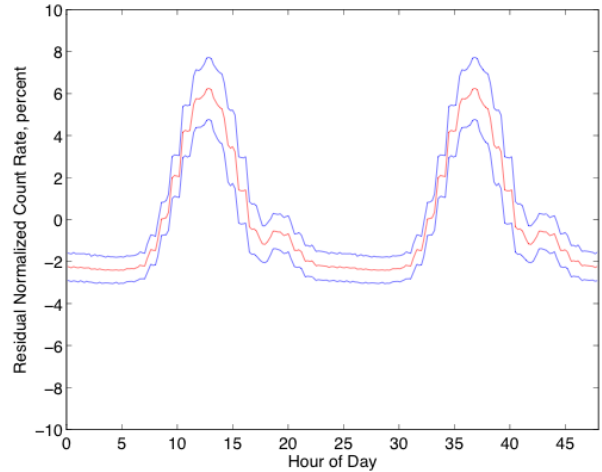

Fig. 5. Residual of normalized gamma measurements, in percent, as a function of hour of day (repeated for clarity). The blue lines indicate the upper and lower offset of 10 times the standard error of the mean.

We next carry out a similar analysis of the distribution of measurements in "hour of year" (dividing the year into 24 equal time intervals). The result is shown in Figure 6. We see that the principal peak occurs at hour of year 11.3, which corresponds to day of year 172 (June 21). This is an interesting result, since it is close to the time of year (June 2) at which dark-matter measurements are expected to have their maximum value (Drukier, Freese, Spergel, 1986), raising the possibility that there may be a cosmic influence on nuclear decay-rates. A further strong peak occurs at hour of the year 23.

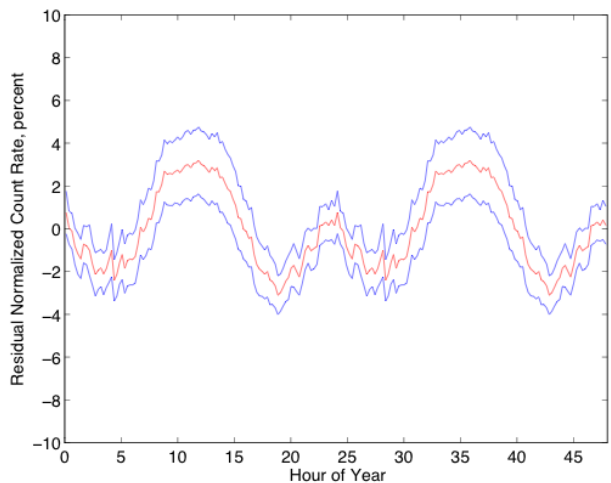

Fig. 6. Residual of normalized gamma measurements, in percent, as a function of hour of year (repeated for clarity). The blue lines indicate the upper and lower offset of 10 times the standard error of the mean.

In view of the strong diurnal variation in the count rate, it is convenient to examine separately measurements made at noon and at midnight. Figure 7 shows 2-hour normalized residuals (in percent) of measurements made at noon (upper panel) and at midnight (lower panel).

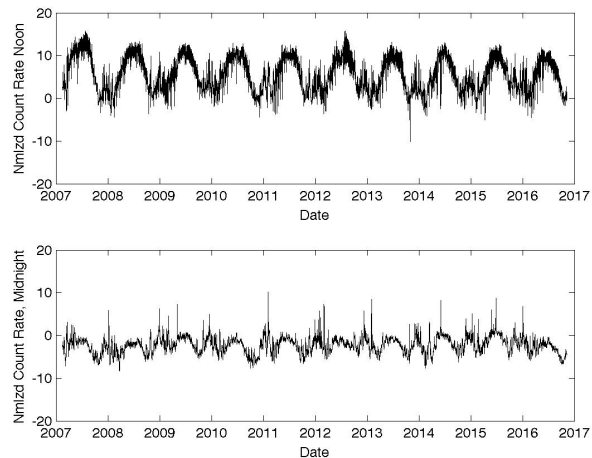

Fig. 7. Residuals in percent of the normalized gamma measurements made at noon (upper panel, hour of day 11.00 to 13.00, mean value 5.66%) and at midnight (lower panel, hour of day 23.00 to 1.00, mean value -2.25%) as a function of date. Note that pulses occur only in the midnight data.

As we expect from Figure 5, the mean value of measurements made at noon is about 8 percent higher than the mean value of measurements made at midnight. In order to attribute these measurements to experimental or environmental influences, it would clearly be necessary to identify either a very strong and complex influence, or two or more such influences.

It is interesting that measurements made at midnight exhibit a number of upward pointing spikes, which we refer to as "pulses", but the same is not true of the noon measurements. Figure 8 shows the fine structure of part of one of the pulses. We see that it extends over a three-day interval, so it seems not to be an experimental artifact.

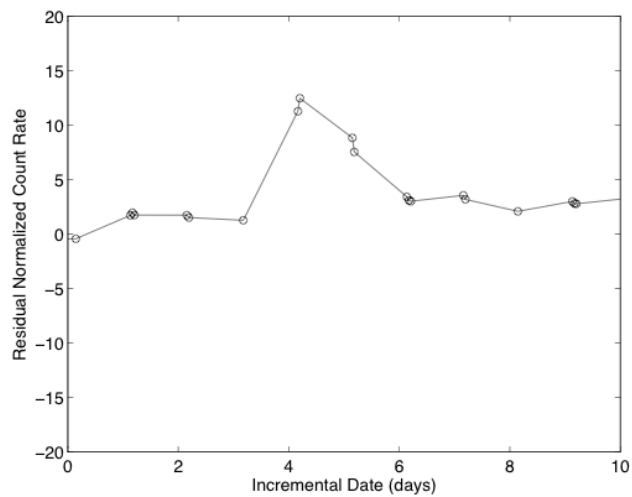

Fig. 8. Expansion of residuals shown in the lower panel of Figure 7, extracting 10 days centered on 2011.092. We see that this pulse has a duration of about 3 days. Data points occur in pairs since the figure shows measurements made at 11.5 pm and 0.5 am. The actual peak probably occurs at about 4.4 days.

## 3 . Power-Spectrum Analysis

We have carried out power-spectrum analyses of the detrended data derived from Equation (3) for the frequency range 0 – 16 year$^{-1}$, using the same likelihood procedure (Sturrock, 2003) that we have used in recent articles (e.g. Sturrock, Fischbach, Scargle, 2016). The power, as a function of frequency, is given by

$$S(\nu) = \frac{1}{2\sigma^2}\sum_r x_r^2 - \frac{1}{2\sigma^2}\sum_r \left(x_r - A e^{i2\pi\nu t_r} - A^* e^{-i2\pi\nu t_r}\right)^2 \qquad (4)$$

where $\sigma$ is the standard deviation of the measurements and, for each frequency, the complex amplitude A is adjusted to maximize S.

We shall be interested in two different types of oscillation in the data: an annual oscillation and harmonics (multiples of the frequency) of that oscillation, and oscillations that we attribute to solar rotation. Since the power of the former is much greater than that of the latter, it is convenient to prepare the relevant figures separately for these two types of oscillation. We therefore prepare figures for the frequency range 0 – 6 year$^{-1}$ for the annual-type oscillations, and for the frequency range 6 – 16 year$^{-1}$ for the rotation-type oscillations.

Figure 9 shows the power spectra for the frequency band 0 – 6 year$^{-1}$ formed from 4-hour bands of measurements centered on noon and midnight. The top 30 peaks in the power spectra are shown in Tables 1 and 2, in which peaks of special interest are shown in bold font. The fact that the powers are so large, and that the spectra formed from noon and midnight data are so different, makes it highly unlikely that the oscillations are due to solely to environmental or experimental factors, but we explore this possibility in Section 5.

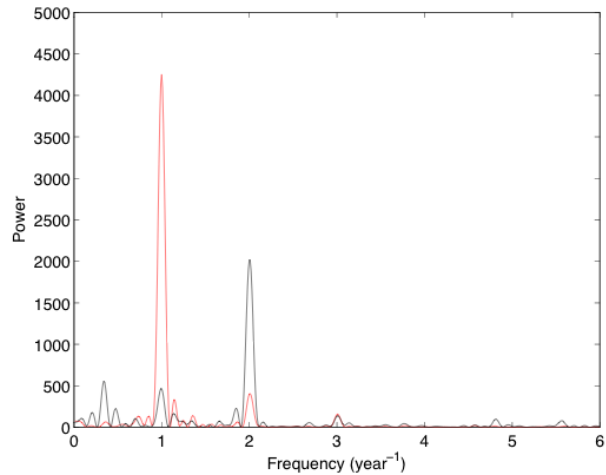

Fig. 9. Power spectra formed from the 4-hour band of measurements centered on noon (red) and midnight (blue) for the frequency band 0 – 6 year$^{-1}$. We see that the biggest daytime oscillation is at 1 year$^{-1}$; the biggest nighttime oscillation is at 2 year$^{-1}$. (cf Tables 1 and 2.)

Table 1. Top 30 peaks in the power spectrum formed from noon data in the frequency band 0 – 6 year$^{-1}$.

| Frequency (year$^{-1}$) | Power | Order |
|---|---|---|
| 0.06 | 71.36 | 9 |
| 0.20 | 19.37 | 18 |
| 0.36 | 64.55 | 10 |
| 0.55 | 36.11 | 13 |
| 0.74 | 133.52 | 7 |
| 0.85 | 137.07 | 6 |
| **1.00** | **4253.63** | **1** |
| 1.15 | 332.13 | 3 |
| 1.25 | 82.10 | 8 |
| 1.36 | 141.28 | 5 |
| 1.47 | 33.07 | 14 |
| 1.57 | 36.31 | 12 |
| 1.67 | 26.46 | 15 |
| 1.77 | 13.62 | 23 |
| 1.86 | 63.64 | 11 |
| **2.01** | **406.39** | **2** |
| 2.14 | 13.34 | 24 |
| 2.38 | 8.18 | 30 |
| 2.54 | 11.81 | 28 |
| 2.68 | 14.14 | 21 |
| 2.88 | 18.69 | 19 |
| **3.01** | **156.77** | **4** |
| 3.14 | 12.29 | 27 |
| 3.23 | 13.93 | 22 |
| 3.44 | 9.85 | 29 |
| 3.55 | 13.33 | 25 |
| 3.78 | 20.42 | 17 |
| 4.57 | 21.78 | 16 |
| 4.80 | 15.97 | 20 |
| 5.60 | 13.17 | 26 |

Table 2. Top 30 peaks in the power spectrum formed from midnight data in the frequency band 0 – 6 year$^{-1}$.

| Frequency (year$^{-1}$) | Power | Order |
|---|---|---|
| 0.09 | 107 | 9 |
| 0.21 | 178.84 | 6 |
| 0.34 | 557.9 | 2 |
| 0.47 | 223.05 | 5 |
| 0.59 | 44.82 | 20 |
| 0.7 | 105.21 | 10 |
| 0.82 | 21.98 | 27 |
| **1.00** | **467.52** | **3** |
| 1.14 | 166 | 7 |
| 1.23 | 74.19 | 14 |
| 1.34 | 76.62 | 13 |
| 1.54 | 32.12 | 21 |
| 1.66 | 73.76 | 15 |
| 1.76 | 27.55 | 24 |
| 1.85 | 230.6 | 4 |
| **2.00** | **2019.66** | **1** |
| 2.16 | 63.7 | 16 |
| 2.68 | 56.15 | 17 |
| 2.88 | 26.59 | 26 |
| **3.01** | **139.04** | **8** |
| 3.14 | 52.87 | 18 |
| 3.55 | 31.16 | 22 |
| 3.77 | 44.92 | 19 |
| **3.97** | **21.1** | **28** |
| 4.45 | 19.83 | 30 |
| 4.58 | 26.89 | 25 |
| 4.81 | 97.37 | 11 |
| **4.95** | **30.7** | **23** |
| 5.57 | 78.37 | 12 |
| 5.83 | 20.27 | 29 |

Figure 10 shows the power spectra for the frequency band 6 – 16 year$^{-1}$ formed from 4-hour bands of measurements centered on noon and midnight. The top 30 peaks in the power spectra are shown in Tables 3 and 4, in which peaks of special interest are shown in bold font. We see that the midnight-centered data show strong evidence of oscillations in the frequency band appropriate for solar rotation, but the noon-centered data show only weak evidence of such oscillations.

Table 3. Top 30 peaks in the power spectrum formed from noon data in the frequency band 6 – 16 year$^{-1}$.

| Frequency (year$^{-1}$) | Power | Order |
|---|---|---|
| 6.07 | 4.41 | 16 |
| 6.54 | 3.34 | 24 |
| 6.72 | 4.49 | 15 |
| 6.86 | 2.54 | 30 |
| 7.31 | 3.26 | 25 |
| **7.45** | **10.68** | **2** |
| 7.81 | 7.79 | 5 |
| 7.96 | 3.49 | 20 |
| 8.17 | 2.61 | 27 |
| **8.47** | **4.10** | **17** |
| 8.85 | 6.49 | 7 |
| 9.21 | 4.63 | 13 |
| 9.38 | 2.83 | 26 |
| **9.48** | **3.44** | **22** |
| **10.31** | **4.99** | **11** |
| 10.74 | 6.40 | 8 |
| 10.90 | 5.90 | 10 |
| **11.34** | **14.86** | **1** |
| 11.69 | 2.57 | 29 |
| 12.37 | 3.69 | 19 |
| **12.65** | **6.75** | **6** |
| 12.86 | 7.93 | 4 |
| 13.13 | 9.56 | 3 |
| 13.29 | 3.45 | 21 |
| **13.67** | **6.02** | **9** |
| 13.89 | 3.35 | 23 |
| 14.14 | 4.86 | 12 |
| 14.99 | 3.73 | 18 |
| 15.24 | 4.52 | 14 |
| 15.65 | 2.58 | 28 |

Table 4. Top 30 peaks in the power spectrum formed from midnight data in the frequency band 6 – 16 year$^{-1}$.

| Frequency (year$^{-1}$) | Power | Order |
|---|---|---|
| 6.13 | 18.51 | 19 |
| 6.49 | 15.76 | 24 |
| 7.18 | 18.85 | 18 |
| 7.32 | 15.75 | 25 |
| **7.45** | **20.66** | **15** |
| 7.80 | 37.13 | 5 |
| 8.30 | 22.19 | 14 |
| **8.46** | **42.38** | **4** |
| 8.71 | 17.02 | 23 |
| 8.87 | 19.64 | 16 |
| 9.21 | 24.77 | 12 |
| **9.44** | **22.55** | **13** |
| 9.95 | 18.20 | 20 |
| 10.24 | 15.33 | 27 |
| 10.76 | 17.13 | 21 |
| 10.93 | 36.42 | 7 |
| **11.35** | **65.47** | **1** |
| 11.70 | 12.61 | 29 |
| 11.91 | 19.07 | 17 |
| **12.35** | **31.73** | **9** |
| 12.50 | 17.08 | 22 |
| **12.63** | **61.35** | **2** |
| 12.86 | 32.15 | 8 |
| 13.13 | 15.71 | 26 |
| **13.67** | **31.13** | **10** |
| 13.90 | 25.37 | 11 |
| 14.14 | 37.11 | 6 |
| 15.00 | 51.32 | 3 |
| 15.23 | 11.31 | 30 |
| 15.70 | 13.73 | 28 |

The strongest peak listed in Table 4 is at 11.35 year$^{-1}$, with a power $S = 65.47$, and the second strongest peak is at 12.63 year$^{-1}$, with a power $S = 61.35$. Such large powers are highly significant. According to the standard interpretation (Scargle, 1982), the probability of finding a peak with power S at a given frequency (interpreting the data approximately as a Gaussian contribution) is given by $P = e^{-S}$ which, for the strongest peak in the rotational band that has $S = 65.47$, gives $P = 3.7 \times 10^{-29}$. We find that there are 37 peaks in the range 10 – 16 year$^{-1}$, so the probability of finding a peak with power 65.47 anywhere in the range 10 – 16 year$^{-1}$ is 1.4 10$^{-27}$.

Since we attach special interest to oscillations with frequencies in the band 10 – 16 year$^{-1}$, which may have their origin in solar rotation, and are less likely than the diurnal or annual oscillations to have their origin in environmental effects, we have used the shuffle test (Bahcall, Press, 1991) to check the validity of assigning the usual probability significance estimate e$^{-S}$ (Scargle, 1982) to peaks. We see, from the result of 10,000 shuffles shown in Figure 11, that this test supports the standard interpretation of the significance of the power as calculated in this article.

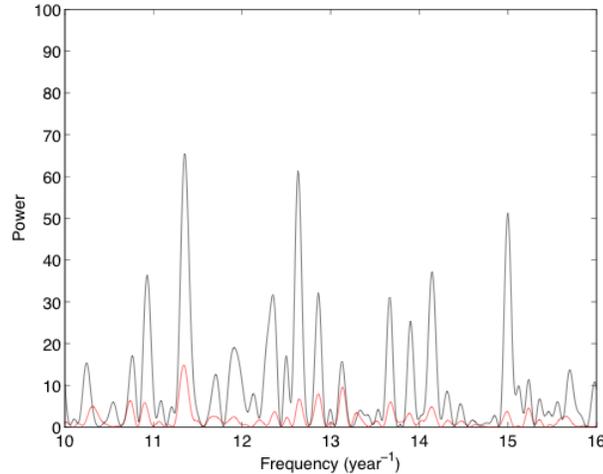

Fig. 10. Power spectra formed from the 4-hour band of measurements centered on noon (red) and on midnight (blue) for the frequency band 10 – 16 year$^{-1}$. We see that there are strong oscillations in the candidate rotational frequency band 10 – 14 year$^{-1}$ (note especially the peaks at 11.35 year$^{-1}$ and 12.64 year$^{-1}$) in the nighttime data, but comparatively small oscillations in the daytime data. (cf Tables 3 and 4.)

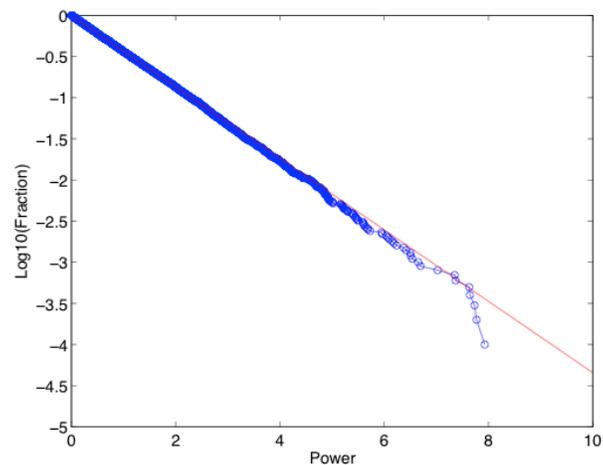

Fig. 11. Result of a shuffle test (with 10,000 shuffles) of the validity of the formula $\exp(-S)$ (shown in red) as an estimate of the significance level of the power S, as calculated by the likelihood procedure used in this section and elsewhere in this article. The blue curve shows, as ordinate, the fraction of the shuffles that have a power larger than the value indicated in the abscissa. We see that the result of the test closely tracks the adopted formula ($P = e^{-S}$).

**4 . Sidereal and Synodic Frequency Pairs**

Schou, Antia, Basu, *et al.*, (1998) have derived an estimate of the internal rotation rate of the Sun as a function of radius from helioseismology data obtained by the Michelson Doppler Imager (MDI). This experiment, which was mounted on NASA's Solar and Heliospheric Observatory (SOHO, in operation from 1997.843 to 2011.282), produced an estimate of the *sidereal* measure of the solar internal rotation rate (as it would be measured from space). Couvidat, Garcia, Turck-Chieze, *et al.* (2003) have analyzed data from the GOLF (Global Oscillations at Low Frequencies) instrument (also on the SOHO spacecraft) and from the LOWL sub-dataset of MDI data. The Coudivat estimate of the solar internal rotation rate is in close agreement with the Schou estimate.

If the rotation axis were normal to the plane of the ecliptic, the rotation rate as measured by an observer on Earth would be not the sidereal rate but the so-called *synodic* rotation rate, which is less than the sidereal rate by 1 year$^{-1}$. Since the axis of rotation of the Sun, as determined from observation of the photosphere, differs from the normal to the ecliptic by only 7 degrees, one would expect that observations of solar rotational phenomena are more likely to exhibit oscillations at synodic frequencies than at sidereal frequencies. It is therefore convenient to present the Schou data in Figure 12 in terms of the presumed *synodic* frequency rather than the sidereal frequency.

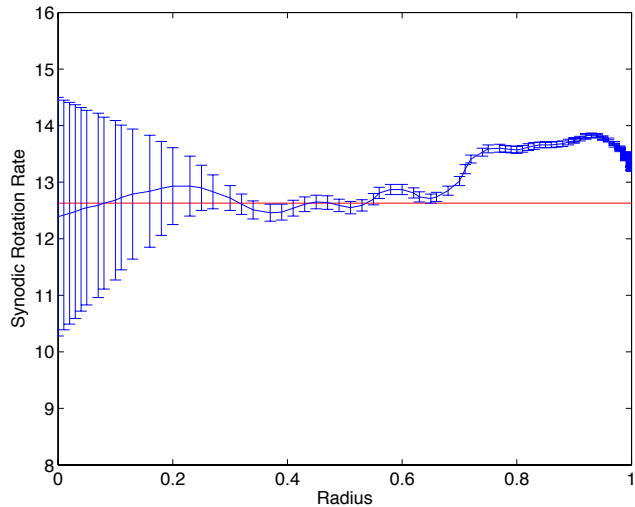

Fig. 12. Estimate (in blue, with error bars) of the synodic solar internal rotation rate as a function of radius, as inferred from helioseismology data acquired by the Michelson Doppler Imager (MDI; Schou, Antia, Basu, et al, 1998). The red line corresponds to a highly significant oscillation at 12.63 year$^{-1}$ in GSI measurements shown in Figure 10 and listed in Table 4. We see that there is an excellent fit to the MDI rotation estimate over the radiative zone (outer radius 0.7, inner radius taken to be 0.3).

We see that the highly significant oscillation noted in Table 4 at 12.63 year$^{-1}$ (the second strongest in the frequency band 6 - 16 year$^{-1}$) is a good match to the synodic rotation rate over the radial range 0.3 to 0.7, which is usually taken to be that of the solar radiative zone. This correspondence suggests that the radon decay process may be influenced by some form of radiation from the deep solar interior (such as the core), and that this radiation is modulated by some process in the solar radiative zone. We shall return to this hypothesis in Section 6.

We have noted that if the rotation axis of some solar region were normal to the plane of the ecliptic, the corresponding rotation rate as measured by an observer on Earth would be the synodic rotation rate. However, if the angle between the solar rotation axis and the normal to the ecliptic is between 0 and $\pi/2$, an observer on Earth will detect *both* the sidereal and synodic rotation rates in a combination that is determined primarily by the orientation of the axis (Sturrock and Bai, 1992). It is therefore interesting to note from Table 4 that oscillations in GSI data include not only the oscillation with frequency 12.63 year$^{-1}$ (power S = 61.35), but also a strong oscillation with frequency 13.67 year$^{-1}$ (S = 31.13). That is to say, *GSI detects oscillations at both the sidereal and synodic rotation frequencies of the radiative zone.*

This fact has two interesting consequences:

( 1 ) The axis of rotation of the radiative zone must depart significantly from the direction of the normal to the ecliptic, and hence departs significantly from the axis of rotation of the photosphere (and presumably of the outer convection zone). In other words, *the radiative zone is an oblique rotator.*

( 2 ) *Neither of the signals at effectively 12.65 year$^{-1}$ and 13.65 year$^{-1}$ can reasonably be attributed to some environmental or experimental artifact, since they clearly form a sidereal-synodic pair of oscillations.*

This raises the question of whether GSI data reveal other sidereal-synodic pairs of frequencies which would presumably be indicative of other obliquely rotating internal regions of the Sun. To examine this possibility, we plot the product of the powers at frequency $\nu$ and at frequency $\nu - 1$, as displayed in Figure 13. We see that there are a number of candidate pairs, suggesting that the solar interior may have quite a complex structure.

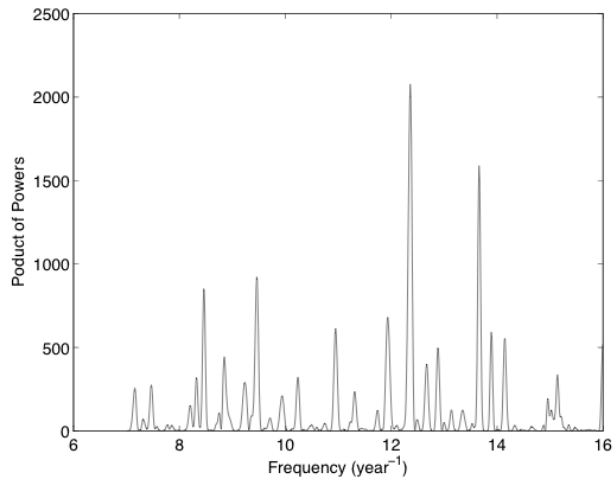

Fig. 13. The product of $S(\nu)$ and $S(\nu - 1)$ over the frequency range $7 - 16$ year$^{-1}$. The four strongest features are found at 12.36 year$^{-1}$, 13.66 year$^{-1}$, 9.46 year$^{-1}$ and 8.46 year$^{-1}$.

The four strongest features in this figure are found at 13.66 year$^{-1}$, 12.36 year$^{-1}$, 9.46 year$^{-1}$ and 8.46 year$^{-1}$. The first feature corresponds to the sidereal-synodic pair that we identified with the radiative zone with sidereal rotation frequency 13.67 year$^{-1}$, as examined by Schou, Antia, Basu, *et al.* (1998) and by Couvidat, Garcia, Turck-Chieze, *et al.* (2003). The second feature is indicative of a similar region with a somewhat lower sidereal rotation frequency, 12.35 year$^{-1}$. We *tentatively* refer to these two regions as an *Outer Radiative Zone* and an *Inner Radiative Zone*, respectively, on the assumption that the region with a rotation rate closest to that of the convection zone will be the one closest to that zone.

The ratio of the powers of the sidereal and synodic frequencies is determined primarily by the angle of inclination of the rotation axis to the normal to the ecliptic (Sturrock and Bai, 1992). We find that this ratio is very similar for both pairs of oscillations, suggesting that the rotation axes of the Inner and Outer Radiative Zones are close to parallel.

The third and fourth features, comprising a triplet with frequencies 9.46 year$^{-1}$, 8.46 year$^{-1}$ and 7.46 year$^{-1}$, appear to be related to a feature with frequency 9.43 year$^{-1}$ in neutrino data acquired by the Super-Kamiokande neutrino observatory (Sturrock and Scargle, 2006). If there is a significant difference between the axis of rotation and the normal to the ecliptic, it is indeed possible for an observer to detect not only the sidereal frequency $\nu$ and the synodic frequency $\nu - 1$ but also the "second lower sideband" frequency $\nu - 2$ (Sturrock and Bai, 1992). In line with our earlier suggestion that the oscillation in Super-Kamiokande measurements is attributable to the solar core, we suggest that 9.46 year$^{-1}$ is the sidereal rotation rate of the core, 8.46 year$^{-1}$ is the synodic rate of the core, and 7.46 year$^{-1}$ is the second lower sideband of the core rotation frequency. We summarize these findings, and our proposed interpretations, in Table 5.

Table 5. Pairs and triplets of peaks separated by 1.00 year$^{-1}$ in the power spectrum formed from midnight data in the frequency band 6 – 16 year$^{-1}$.

| Approximate Frequency (year$^{-1}$) | Power | Proposed Interpretation |
| --- | --- | --- |
| 7.45 | 20.66 | |
| 8.45 | 42.38 | Core |
| 9.45 | 22.55 | |
| | | |
| 11.35 | 65.47 | Inner Radiative Zone |
| 12.35 | 31.73 | |
| | | |
| 12.65 | 61.35 | Outer Radiative Zone |
| 13.65 | 31.13 | |

## 5 . Hour-of-Day-Frequency Spectrogram Analyses

We may obtain additional information about the oscillations by examining the power spectrum as a function of some other variable. For instance, we can examine the power as a function of both frequency and date of sampling, as in our recent analysis of BNL and Super-Kamiokande data (Sturrock, Fischbach, Scargle, 2016). However, we see in Figures 9 and 10 that the power spectrum varies very strongly with hour of day. For this reason, it proves more informative to form hour-of-day-frequency spectrograms, in which we examine the frequency spectrum as a function of hour of day.

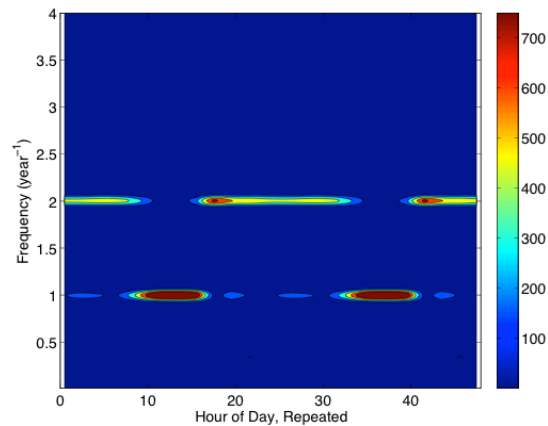

Figure 14. Power spectrum as a function of hour of day for the frequency range 0 – 4 year$^{-1}$. Note that the dominant feature is an "island" centered on noon at 1 year$^{-1}$.

Figure 14 displays such a spectrogram for the frequency range 0 – 4 year$^{-1}$. We see that there is a very strong signal at 1 year$^{-1}$ that extends approximately over the range 9 – 15 hours, with a peak at noon. The second strongest signal is at 2 year$^{-1}$ and extends approximately over the range 16 hours to 7 hours of the next day, with a peak at 18 hours. These results are consistent with Figure 9 and with Tables 1 and 2.

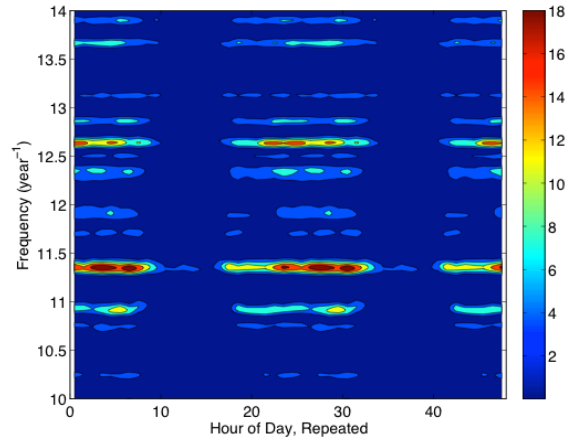

Figure 15. Power spectrum as a function of hour of day for the frequency range 10 – 14 year$^{-1}$. Note that the dominant features are "islands" at frequencies 11.4 year$^{-1}$ and 12.6 year$^{-1}$ centered near 3 am.

Figure 15, which displays the power for the frequency range 10 – 14 year$^{-1}$, shows two strong signals, one at approximately 11.35 year$^{-1}$, and the other at approximately 12.65 year$^{-1}$. It should be noted that *these frequencies are prominent in power spectra and spectrograms formed from both BNL and Super-Kamiokande data (Sturrock, Fischbach and Scargle, 2016)*. In the GSI data here investigated, these are both nighttime features extending from 20 hours to 8 hours the next day, with peaks near midnight. These two features correspond to the strongest and second-strongest peaks in the nighttime power spectrum shown in Figure 10 and listed in Table 4.

In order to assess the possible significance of experimental and environmental influences on the GSI measurements, it is interesting to compare these spectrograms formed from GSI data with spectrograms formed from measurements of environmental parameters that were recorded as part of the same experiment. Figure 16 is a display in which we compare the spectrogram formed from gamma measurements for the frequency range 0 – 4 year$^{-1}$ with spectrograms formed from ambient temperature, ambient pressure, and battery voltage measurements for the same frequency range.

Apart from the expected annual oscillation, the three environmental spectrograms are almost featureless. The temperature spectrogram shows hardly any variation, and the pressure and voltage spectrograms show only slight enhancements near midnight and noon, respectively. None of the three replicates the "island" structure at 1 year$^{-1}$ or the more extended feature at 2 year$^{-1}$ that are so pronounced in the gamma spectrogram. Hence this comparison provides no support for the conjecture that variability in the gamma measurements is attributable to environmental influences.

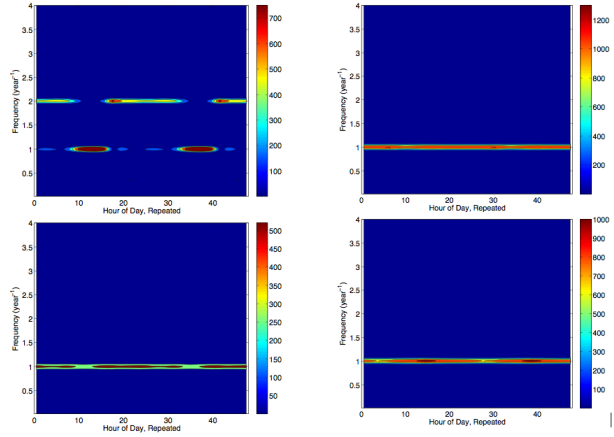

Figure 16. Spectrograms for hour of day and for the frequency range 0 – 4 year$^{-1}$ for gamma-ray measurements (upper left panel), ambient temperature (upper right panel), ambient pressure (lower left panel), and voltage supply (lower right panel). We see that the dominant feature in the gamma-ray measurements is an "island" centered on noon and 1 year$^{-1}$. There is no corresponding "island" in any of the plots made from the environmental measurements, indicating that the annual oscillation in the gamma-ray measurements cannot be attributed to environmental influences.

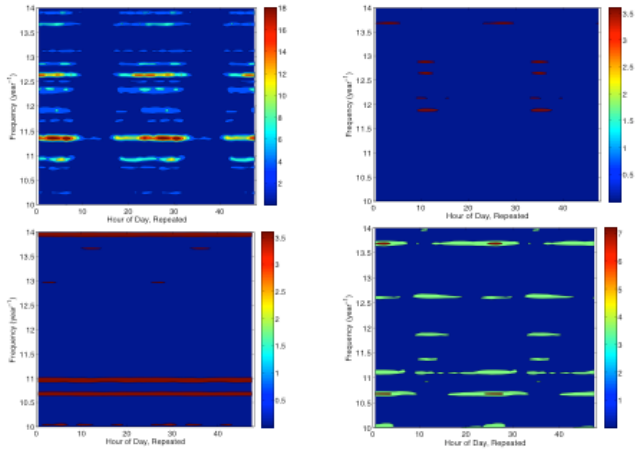

Figure 17. Spectrograms for hour of day and for the frequency range 10 – 14 year$^{-1}$ for gamma-ray measurements (upper left panel), ambient temperature (upper right panel), ambient pressure (lower left panel), and voltage supply (lower right panel), all to same color scale. The pattern of the gamma-ray plot does not resemble any of the environmental plots. Furthermore, the environmental plots have smaller powers than the gamma-ray plots, which is the opposite of what one would expect if the gamma-ray modulation were caused by the environmental modulations.

Figure 17, which examines the same spectrograms for the frequency range 10 – 14 year$^{-1}$, is also interesting. These spectrograms have all been formed with the same colorbar, from which it is clear that the oscillations in the environmental measurements are significantly *weaker* than the oscillations in the gamma measurements, which is the opposite of what one would expect if oscillations in the gamma measurements responded significantly to environmental influences. This fact clearly indicates that oscillations in the decay measurements are highly unlikely to be attributable to environmental influences such as temperature, pressure or voltage. Furthermore, none of the environmental spectrograms reproduces the strong rotational modulations evident in the gamma spectrogram.

## 6 . Discussion

We first point out that patterns evident in our current investigation are fully consistent with patterns found in our earlier analysis of the first three years of GSI data (Sturrock, Steinitz, Fischbach, *et al.,* 2012). For instance, spectrograms shown in Figures 14 and 15 are fully consistent with Figures 11 and 13 of our 2012 article. However, as would be expected, our current results are statistically more significant than our earlier results.

Our current results are also fully consistent with the results of our recent analysis of BNL and Super-Kamiokande data (Sturrock, Fischbach, Scargle, 2016). Figure 10 of our 2016 article shows that a spectrogram formed from $^{36}$Cl data shows evidence of oscillations at about 11.3 year$^{-1}$ and 12.6 year$^{-1}$, which correspond to the two strongest features in our current Figure 15. An earlier article (Sturrock and Scargle, 2006) and Figure 14 of our 2016 article show that an oscillation near 9.5 year$^{-1}$ is a prominent feature of Super-Kamiokande data; the same oscillation is found in GSI data, as listed in Tables 4 and 5.

In the Introduction (Section 1), we commented on various unsatisfactory attempts to explain the variability of decay rates in terms of environmental effects. We can now add that, as exemplified by Figures 16 and 17, the annual and rotational oscillations in GSI measurements cannot be attributed to known environmental influences.

Another point we wish to stress is that it would not be enough to identify a single oscillation as being potentially of experimental origin. We now find that *each of the oscillations at 11.35 year$^{-1}$ and 12.65 year$^{-1}$ is one member of a sidereal-synodic pair of frequencies.* We are not aware of any environmental effect that might lead to a pair of oscillations separated in frequency by 1 year$^{-1}$. Nor can such pairs be attributed to aliasing resulting from some unknown nonlinear process in the experiment, since aliasing generates pairs of oscillations with frequencies both above and below the parent frequency, and there is no evidence of significant oscillations at either 10.35 year$^{-1}$ or 11.65 year$^{-1}$.

It is interesting to recall that analysis of Homestake solar-neutrino data yields several oscillations with frequencies separated by 1 year$^{-1}$ (Sturrock, Walther, Wheatland, 1997). This strengthens the case that the generation of oscillations separated in frequency by 1 year$^{-1}$ is an intrinsic property of whatever mechanism is responsible for oscillations in nuclear decay rates. Internal rotation is the leading candidate for such a role.

Perhaps the most striking discovery of our analysis of GSI data (which was anticipated by our 2012 article) is the dramatic difference between daytime measurements and nighttime measurements, as exemplified by Figures 5 and 7. We find strong evidence of a solar influence (in terms of rotational oscillations) in nighttime data, but little evidence in daytime data. This difference is due, we propose, to the geometry of the experiment, which is shown in Figure 1. There is a significant separation between the source of decay products (in the lower part of the experiment) and the detectors of those products (at the top). The gamma detector will therefore be more responsive to gammas traveling upward than to gammas traveling downward. As a result, the experiment primarily detects a solar stimulus that has traveled through the Earth, a fact that clearly implicates neutrinos.

We propose, as a hypothesis to be tested, that

*Neutrinos can stimulate beta decays* (possibly as part of a chain that may include also alpha decays) *and that, when this occurs, the secondary products of the decay tend to travel in the same direction as the stimulating neutrino.*

To be cautious, one might also consider the possibility that neutrinos may stimulate alpha decays, but – to the best of our knowledge – there is no evidence of variability in nuclear processes that involve only alpha decays. (See, for instance, Parkhomov, 2010a, 2010b, 2011.) Our current analysis of GSI data is restricted to measurements of gamma radiation that has its origin in beta decays, not in alpha decays. For these reasons, we here confine our hypothesis to a possible influence of neutrinos on beta decays. If future experiments yield evidence of an *intrinsic* variability of alpha decays, it will be necessary to revise this hypothesis.

These results have implications concerning solar structure. The prominent oscillations near 11.4 year$^{-1}$ and 12.7 year$^{-1}$ suggest that the radiative zone may perhaps be divided into two regions, which we refer to as an "inner radiative zone" and an "outer radiative zone," as indicated schematically in Figure 18. However, Figure 13 may alternatively suggest that the radiative zone comprises many layers, or perhaps that (like the convection zone) it may have a quasi-turbulent structure that is continually changing. This would be consistent with Figures 6 through 11 of our 2016 article (Sturrock, Fischbach, Scargle, 2016), which show evidence of variability in $^{36}$Cl and $^{32}$Si measurements. These options could conceivably be evaluated by new helioseismology experiments or by re-analyses of existing helioseismology data.

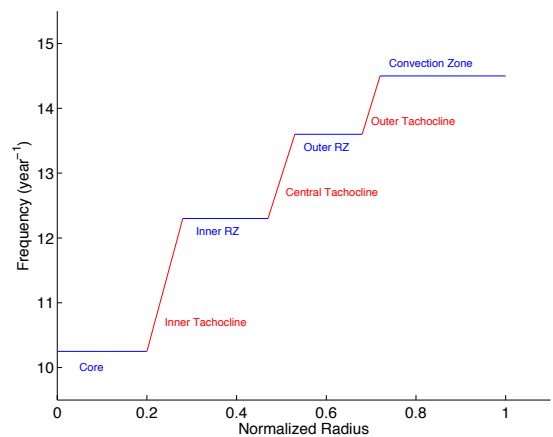

Figure 18. Schematic diagram of inferred solar rotation. The frequencies are accurately inferred from the GSI measurements, but the radial locations are unknown except for the location of the Convection Zone, the Outer Tachocline, and the outer limit of the Outer Radiative Zone.

If the association of neutrino and beta-decay oscillations with solar rotation proves valid, one will need to understand theoretically the mechanism that leads to this association. One promising theoretical approach seems to be the RSFP (Resonant Spin Flavor Precession) mechanism by which neutrinos of one flavor, traveling through a plasma permeated by magnetic field, can change to a different flavor (Akhmedov, 1958, 1998; Pulido, Das, Picariello, 2010; Sturrock, Fischbach, Javorsek, *et al.,* 2013; Sturrock, Fischbach, 2015).

The "pulses" that are evident in the lower panel of Figure 7 are intriguing. We have been unable to identify any similar features in any record of known solar phenomena. This raises the possibility that they may have their origin in cosmic neutrinos that have passed through or near the Sun and are traveling away from the Sun.

Resolution of the question of variability of nuclear decay rates might be advanced by a new generation of experiments that are based on the GSI model but have enhanced energy and directional sensitivity, such as could be obtained with a geometrical array of energy-sensitive detectors. Such experiments may help resolve the differences between observations of gamma photons from radon-chain decays in the GSI experiment and in the related but different experiment of Bellotti, Broggini, Di Carlo, *et al.* (2015). Some experiments may lead to a more complete understanding of radon decays that would be relevant to geology, geophysics, nuclear physics, solar physics, and possibly astrophysics.

If the influence of neutrinos on radioactive material is as large as estimated in Appendix A, this influence may be open to experimental tests of force and torque as suggested in Sturrock. Fischbach, Javorsek *et al.* (2013).

**Acknowledgments**
The Geological Survey of Israel supports the operation of EXP #1. Uri Malik helps with the technical aspects and Oksana Piatibratova participates in the organization and processing of the data. We thank Jeff Scargle for perceptive and enlightening comments.

**Appendix A. Effective cross section**
We assume that the main solar influence on $^{222}$Rn is due to the most abundant solar
neutrinos, i.e. pp neutrinos, which have a mean energy of about 300 keV. According to Bahcall (1989), the flux at Earth is estimated to be $10^{11.4} \text{cm}^{-2}\text{s}^{-1} \text{MeV}^{-1}$. Adopting a mean energy of 300 keV, this gives the following pp-neutrino flux:

$$F_\nu = 10^{10.9} \text{cm}^{-2}\text{s}^{-1}. \tag{A.1}$$

Power-spectrum analysis of the Super-Kamiokande measurements (Sturrock, Scargle, 2006) gives the following estimate of the depth of modulation of the solar neutrino flux at 9.43 year$^{-1}$:

$$\Delta_\nu = 10^{-1.3}. \tag{A.2}$$

Hence the amplitude of the sinusoidal oscillation of the neutrino flux is given by

$$\Delta F_\nu = \Delta_\nu \times F_\nu = 10^{9.6} \text{cm}^{-2}\text{s}^{-1}. \tag{A.3}$$

The half life of $^{222}$Rn is 3.8 days, which leads to the decay rate

$$\Gamma = \frac{\ln(1/2)}{10^{5.5}} = 10^{-5.7} \text{ s}^{-1} . \tag{A.4}$$

The depth of modulation of the 9.43 year$^{-1}$ oscillation in the radon count rate is found to be

$$DOM(\Gamma) = 0.0015 = 10^{-2.8} . \tag{A.5}$$

Hence the amplitude of the 9.43 year$^{-1}$ oscillation in the decay rate is given by

$$\Delta\Gamma = DOM(\Gamma) \times \Gamma = 10^{-8.5} \text{s}^{-1} . \tag{A.6}$$

We define an effective cross section for the influence of the pp solar neutrino flux on the $^{222}$Rn decay rate by

$$\Delta\Gamma = \sigma(^{222}\text{Rn,pp}) \times \Delta F_\nu . \tag{A.7}$$

On using the estimates given in equations (3) and (6), we obtain

$$\sigma(^{222}\text{Rn,pp}) = 10^{-18.3} \text{ cm}^2 . \tag{A.8}$$

This is close to the cross-section ($10^{-18.4}$ cm$^2$) recently estimated for the influence of $^8$B neutrinos on the beta-decay rate of $^{32}$Si, but substantially larger than that ($10^{-21.6}$ cm$^2$) estimated for the influence of $^8$B neutrinos on the beta-decay rate of $^{36}$Cl (Sturrock, Fischbach, 2015).